\documentclass[vecphys]{svmult}

\usepackage{makeidx}         
\usepackage{graphicx}        
\usepackage{multicol}        
\usepackage[bottom]{footmisc}

\makeindex             

\begin{document}

\title*{High-Speed Optical Spectroscopy}
\author{T.R. Marsh\inst{1}}
\institute{Department of Physics, University of Warwick Coventry CV4 7AL, UK
\texttt{tom.marsh@warwick.ac.uk}}

%
%

\maketitle

\begin{abstract}
The large surveys and sensitive instruments of modern astronomy are
turning ever more examples of variable objects, many of which are
extending the parameter space to testing theories of stellar evolution
and accretion. Future projects such as the Laser Interferometer Space Antenna
(\emph{LISA}) and the Large Synoptic Survey Telescope (\emph{LSST}) will only add
more challenging candidates to this list. Understanding such objects
often requires fast spectroscopy, but the trend for ever larger detectors
makes this difficult. In this contribution I outline the science made
possible by high-speed spectroscopy, and consider how a combination of
the well-known progress in computer technology combined with recent
advances in CCD detectors may finally enable it to become 
a standard tool of astrophysics.
\end{abstract}

\section{Introduction}
\label{sec:intro}

High-speed photometry has a long history in optical astronomy. The
late 1960s and early 1970s saw the combination of digital recording
techniques and photomultiplier tubes to give photon counting
high-speed photometry, e.g. \cite{Nather:HighSpeed}. Equivalent
high-speed spectroscopy has never been as straightforward and is still
not a standard technique.  There are several reasons for this. First,
it is technically more difficult and expensive to record spectra at
high-speed as one needs fast 2D imagers rather than single pixel
devices. An early example, which was developed for faint object
spectroscopy but could also take fast spectra, was the Image Photon
Counting System (IPCS) \cite{Boksenberg:IPCS}. The IPCS illustrates a
problem typical of high-speed spectroscopy, because although it could
take spectra with exposure times well below 1~second, it was rarely
used to do so because its design limited its maximum count rate in any
given pixel to less than 1~photon/second. A second difficulty of high-speed 
spectroscopy in the era of the IPCS was simply that it could produce more data than
the computer technology of the 1970s and early 1980s could easily deal
with; the author of this chapter recalls an observing run on the
Anglo-Australian telescope which produced enough magnetic tape to run
to and from the local town Coonabarabran 40 km away twice over.
Computer technology has since taken enormous strides of course, but the
1980s brought the biggest obstacle of all for high-speed spectroscopy,
namely Charge-Coupled Devices (CCDs). Although wonderful detectors in
many respects, the CCDs employed in most observatories have two
significant disadvantages: they are slow, often taking several tens of
seconds to read out, and significant noise is added to each pixel,
burying the small signals characteristic of high-speed spectroscopy.
The slow readouts of standard CCDs has lead to a small revival of the old
photographic method of ``trailed spectra'' where the target is moved along the
slit to give time-resolved spectra \cite{Falter:pg1605,Schwope:HUAqr1}, but
the disadvantages of this method, such as seeing-dependent time-resolution, mean
that it is really only a stop gap measure and I will not consider it further.

Despite the problems of CCDs, their other excellent characteristics,
in particular their high quantum efficiency (QE), have lead to their complete
dominance of optical astronomy, and so the focus of this chapter will be
tilted towards the use of CCDs for high-speed spectroscopy. Moreover, CCDs are
so dominant that they provide us with an empirical definition of what
``high-speed'' means when applied to spectroscopy, i.e.  any spectroscopic
observation that is difficult to carry out with normal CCDs is ``high-speed''.
This does not always mean very fast: readout noise can make echelle
spectroscopy of an 18$^\mathrm{th}$ magnitude object difficult to carry out
with a time resolution shorter than 10 minutes.

With the above broad definition in mind, an overview is given of the science
that high-speed spectroscopy makes possible followed by the technical
requirements and the possible application of electron-multiplying CCDs to this area.

\section{Scientific Motivation}
\label{sec:science}
As the introduction hinted, for a variety of reasons, high-speed spectroscopy
has yet to take off fully and many applications exist only in the imagination.
There are nonetheless a reasonable number of published examples which give an
idea of what to expect, and in addition we have the many applications of
high-speed photometry to draw on as a resource when considering what can be
learned from high-speed spectroscopy. In this section I will look at what
high-speed spectra can tell us when applied to the following phenomena,
sticking as far as possible to what is already known:
\begin{itemize}
\item pulsating white dwarfs and subdwarf B (sdB) stars
\item accreting binary stars
\item white dwarf binary stars 
\end{itemize}
Before doing so, it is worth mentioning objects yet to be discovered because
surveys are being planned to look for short-timescale variable phenomena which
are bound to discover many objects that will need high-speed spectroscopy.
Pre-eminent perhaps, although still far-off, is the Large Synoptic Survey
Telescope (\emph{LSST}), in which it is planned to use an $8.4\,$m telescope
to survey the observable sky once every 3 nights in 15-second exposures.

\subsection{Pulsating white dwarfs and sdB stars}
\label{subsec:pulsations}
As well demonstrated by helioseismology, pulsations in stars can give us unique
insights into stellar structure. The area of asteroseismology tries to use
pulsations in other stars in this manner. It is observationally demanding first
and foremost because of the need for long, uninterrupted time series in order
to obtain clean time series. In the case of white dwarfs and sdB stars,
which have pulsation periods of order 100 to 1000 seconds, fairly fast
observations are also a requirement. If all goes well, it is possible to pin
down stellar parameters with great precision. For instance studies of the pulsating
sdB stars PG0014+067, PG1219+534 and Feige~48 have measured their surface gravities to
$\sim 1$\%, their masses to $\sim 2$\% and also determined the masses of their
hydrogen envelopes, for which no other method exists, to $\sim 10$\%
\cite{Brassard:PG0014,Charpinet:PG1219+534,Charpinet:Feige48}. Despite this,
even large datasets can leave stars unsolved, especially if they display few
pulsation modes, the difficulty being the secure identification of a particular
eigenmode with a given frequency.

\begin{figure}
\centering
\includegraphics[width=0.8\textwidth]{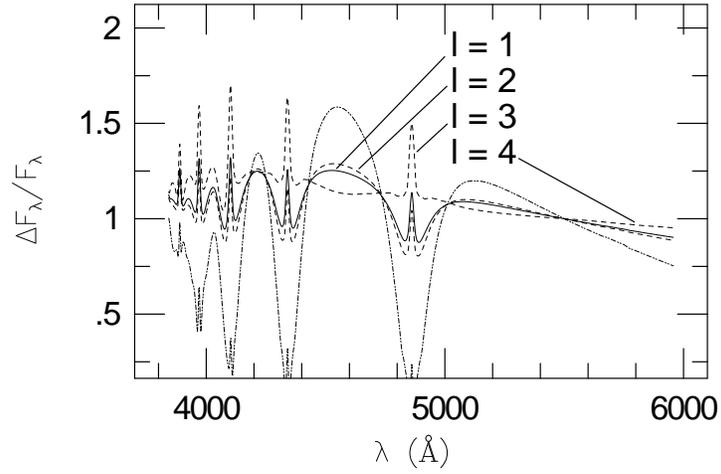}
\caption{The pulsation amplitudes versus wavelength for spherical harmonics
$Y_{lm}$ with different $l$-values. Each curve is normalised by its value at 
5500\AA. The figure is Fig.~3 taken from \protect\cite{Clemens:G29-38}}
\label{fig:clemens3}
\end{figure}

\begin{figure}
\centering
\includegraphics[width=0.8\textwidth]{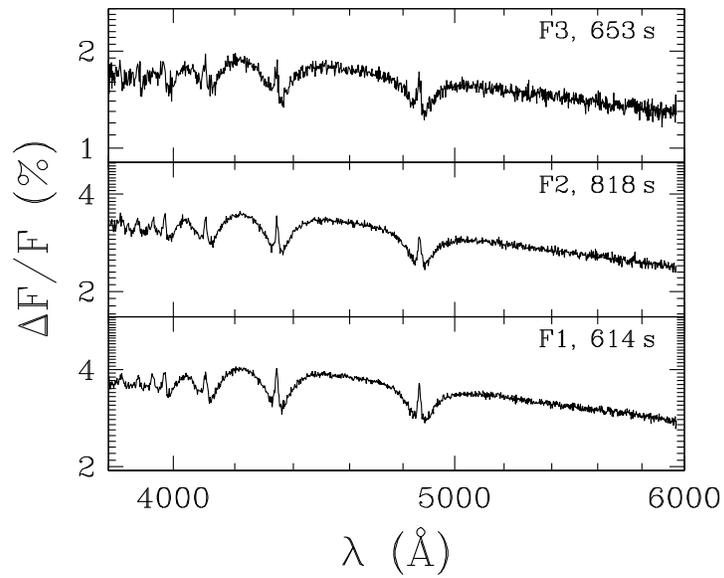}
\caption{Spectra of the first pulsation amplitudes of the ZZ~Ceti star G29-38
observed on Keck-II/LRIS by \cite{vanKerkwijk:G29-38}.
The figure is adapted from Fig.~4 of \protect\cite{Clemens:G29-38}}
\label{fig:clemens4}
\end{figure}

It is unlikely that spectroscopic data can ever be taken with the same time
coverage and uniformity as the photometric data. However spectra can add extra
information missing from the photometric studies because the variation of limb
darkening with wavelength in combination with the different patterns of
different eigenmodes can lead to different variations of pulsation amplitude
with wavelength \cite{Robinson:G117-B15A}. Figure~\ref{fig:clemens3} shows
predictions \cite{Clemens:G29-38} of the amplitudes of different eigenmodes for the 
ZZ~Ceti star G29-38. This shows that spectra may allow $l=3$ and $l=4$
eigenmodes to be  distinguished from $l=1$ and $l=2$ eigenmodes. A
signal-to-noise ratio of 1000 only gives a signal-to-noise of 10 in the
amplitude spectrum when the amplitude itself is of order one percent as it is
in ZZ~Ceti stars, and therefore extremely high quality data are needed
to separate $l=1$ from $l=2$. Thus this application requires moderately fast 
data acquisition to resolve the pulsations, but also a large aperture to
give high signal-to-noise. Figure~\ref{fig:clemens4} shows the result of
4~hours of Keck~II/LRIS time devoted to the brightest ZZ~Ceti star, G29-38
($V = 13$) \cite{vanKerkwijk:G29-38,Clemens:G29-38}. The mean spectrum of these
data appears almost noiseless \cite{Clemens:G29-38}, as might be expected
for a bright star on a large telescope, but comparing Figs.~\ref{fig:clemens3} 
and \ref{fig:clemens4} shows how necessary the large aperture is in this case.

This case illustrates another important feature of high-speed spectroscopy: it
does not have to be especially ``high-speed'' by photometric standards to strain
the capabilities of standard instrumentation: the data were taken
as $700 \times 12\,$second exposures, but for each exposure there were a
further 12~seconds ``deadtime'' for readout, etc., so the observations were only 50\%
efficient. This is quite common; indeed 12~seconds is commendably short in this
respect. Reducing deadtime is probably the single most effective way of enabling
high-speed spectroscopy and would save countless hours of precious telescope time.
 
The end result of the work shown in Figs.~\ref{fig:clemens3} and
\ref{fig:clemens4} is that most of the modes are $l = 1$ modes.  This shows the 
power of spectroscopy because in principle this method can be applied to stars
showing just a few modes where it would certainly not be possible to unravel the
modes purely photometrically.

\subsection{Accreting binary stars}
\label{subsec:accretion}

I now move on to systems which display variability on much shorter timescales
than the white dwarf and sdB pulsators. The shortest dynamical timescales in
accreting binaries with compact objects range from $1$ to $10$ seconds in the case
of accreting white dwarfs down to about a millisecond in the case of the neutron
stars and stellar mass black holes. Variability on seconds timescales is well
established from the ``dwarf nova oscillations'' displayed by dwarf novae during
their outbursts. To my knowledge, dwarf nova oscillations have only been
observed spectroscopically with sufficient speed to detect them once
\cite{Steeghs:V2051Oph}, but this single observation was enough to show the
great diagnostic potential spectra may hold for these poorly-understood but
remarkable phenomena.

The variability timescales seen in black-hole and neutron star systems can
depend upon the system brightness and instrumental limitations as much as the
object. X-ray variability at kilohertz frequencies has been seen in X-ray
binaries, while optical variability on timescales well below $0.1$ seconds has
been seen in bright systems \cite{Motch:GX339-4}. Using ULTRACAM on the Very
Large Telescope (VLT) we
found significant flaring on timescales of a few seconds in a faint quiescent
black-hole accretor (Shahbaz et al, in prep). One can only speculate upon the
spectroscopic signatures of this variability at optical wavelengths as no such
observations have been made, but it is very likely that there will be some,
although in the quiescent black-hole case only an ``Extremely Large
Telescope'' (ELT) would be up to the task. A case of particular interest is
the fluorescent Bowen blend emission from the donor stars seen in some X-ray
binaries and which in some cases has revealed the donor star for the very
first time \cite{Steeghs:ScoX-1}.  This emission is presumably X-ray driven,
and can be expected to respond to X-ray variability; some evidence of this has
been obtained using narrow-band photometry with ULTRACAM \cite{Casares:echo},
but a much cleaner signature could come from high-speed spectra.
 
\begin{figure}
\centering
\includegraphics[width=0.8\textwidth]{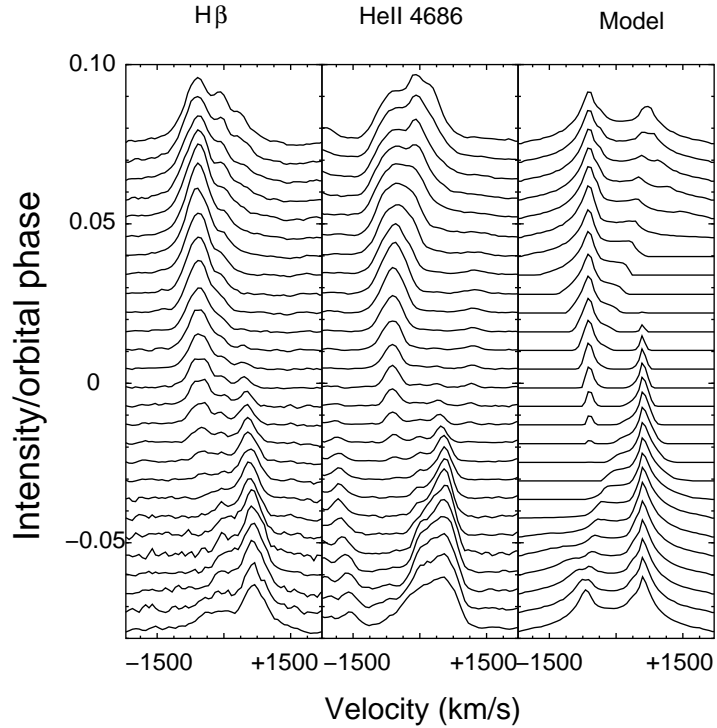}
\caption{The first and second panels show two of the emission lines of the dwarf nova
  IP~Peg  as they are eclipsed, with orbital phase increasing up the plot. The spectra were taken
on Subaru during an outburst of IP~Peg; the lines are dominated by light from
the bright accretion disc during this state. The second panel shows a
simple computation of the eclipse of a disc in prograde, Keplerian rotation.
Figure adapted from Fig.~2 of \protect\cite{Ishioka:IPPeg}}
\label{fig:ishioka2}
\end{figure}

A dramatic example of spectral variability from an accreting binary is shown in
Fig.~\ref{fig:ishioka2}.
This shows the eclipse phases of the white dwarf accretor IP~Peg in one of its
temporary high states (outburst) \cite{Ishioka:IPPeg}. The lines from this
system come primarily from the accretion disc and have the well-known
double-peaked profiles that come from Doppler shifting in accretion discs
\cite{Horne:Emlines}.  The spectra plotted cover about 30 minutes and show
very large changes as first the approaching side of the disc is obscured,
affecting the blue-shifted parts of the lines, followed soon after by their
red-shifted counterparts. The right-most panel of Fig.~\ref{fig:ishioka2}
shows a simple model which captures much of the phenomenology of the data
while at the same time showing significant discrepancies. The nature of these
is not understood although it is possible that the disc during outburst is far
from Keplerian in nature. This example is an even starker indication of the
problems often faced with facility instruments: IP~Peg reaches $V = 12$ in
outburst, yet despite using the $8\,$m Subaru, the time resolution here was a
sluggish 80 seconds, each spectrum consisting of 30 seconds exposure followed
by 50 seconds of deadtime. The telescope, instrument and object would have
allowed \emph{much} higher speed than this. Ideally one would want to sample fast
enough to resolve structure comparable in size to the white dwarf, which would
be about 5 seconds in this case. Unfortunately the detector/data acquisition
system was not up to the job, a not-uncommon situation, as instruments are
rarely built with high-speed applications in mind.

I finish off this section with a look at Doppler imaging as applied to
accreting binary stars in the form of Doppler tomography \cite{Marsh:Doppler,
Marsh:doppler_review}. Doppler tomography uses the information in line profiles
as a function of phase to image binaries. Notable successes have been the
discovery of spiral structure in accretion discs during outburst
\cite{Steeghs:Spiral,Groot:UGem}
and the unravelling of the complex accretion structures in the magnetic
polar class of accreting white dwarfs \cite{Schwope:HUAqr1}. Doppler
tomography provides a quantitative illustration of the need for high-speed
spectroscopy and is useful in defining requirements for such work. This is
because the resolution in Doppler tomography is limited by \emph{both} spectral
\emph{and} time resolution. If one is aiming for a spatial resolution $\Delta
x$, then this imposes the following restrictions on the spectral resolution $R_\lambda =
\lambda/\Delta \lambda$ and the time resolution $\Delta t$:
\begin{eqnarray}
R_\lambda & \sim & (c/\Omega) (\Delta x)^{-1}, \label{eq:R}\\
\Delta t & \sim & V^{-1} \Delta x,\label{eq:delta}
\end{eqnarray}
where the orbital angular frequency $\Omega = 2\pi/P$ where $P$ is the orbital
period, and $V$ is the velocity
of the feature being imaged. Small $\Delta x$ requires large $R_\lambda$ and
small $\Delta t$.  These relations can be combined into a single one relating
spectral and time resolution:
\begin{equation}
\Delta t \sim \Omega^{-1} \frac{c}{V} R_\lambda^{-1}.
\end{equation}
Taking typical values $P = 1.5\,$hours and $V = 700
\,\mathrm{km}\,\mathrm{s}^{-1}$, and assuming that we are working on a
spectrograph with $R_\lambda = 10$,$000$, then we require $\Delta t < 30\,$seconds
in order that smearing during the exposures does not degrade the resolution.
Equations~\ref{eq:R} and \ref{eq:delta} imply that the number of counts per
detector resolution element per
exposure scales as $(\Delta x)^2$, and thus Doppler tomography can become
challenging even on quite bright objects and large telescopes.

\subsection{White dwarf binary stars}
\label{subsec:whitedwarf}

\begin{figure}
\centering
\includegraphics[width=0.8\textwidth]{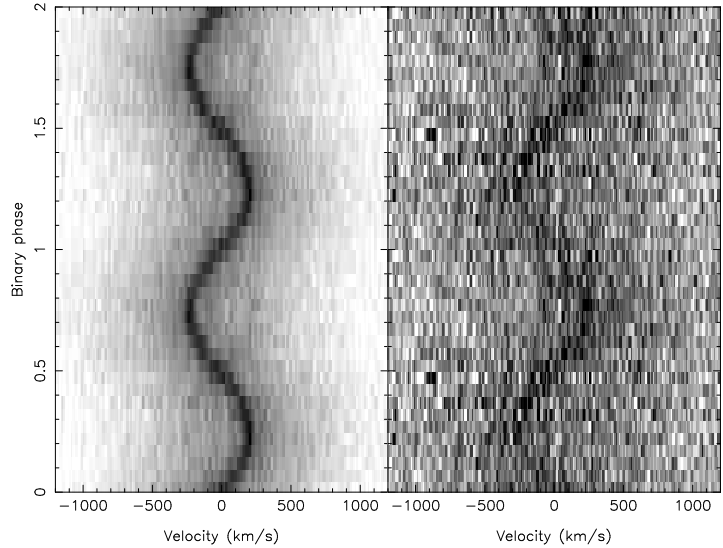}
\caption{Trailed spectra around H$\alpha$ of the shortest period double white dwarf, WD0957-666,
  based upon the data of \protect\cite{Maxted:DDmassratios}. In the left panel,
  the brightest of the two white dwarfs is clearly seen; the right-hand panel
  shows the result of subtracting a fit to the brightest star which reveals its
  companion. The two stars will merge under gravitational radiation in about 200
  million years}
\label{fig:wd0957-666}
\end{figure}

For my final example of applications of high-speed spectroscopy I turn to binary
stars with white dwarf components. This is another case where there are only a
few examples, but where one can point to future applications which will prove tough
for even the largest telescopes. Work over the past decade has
established the existence of a huge population of detached, close double white
dwarfs, with orbital periods of a few days or less
\cite{Marsh:friends,Napiwotzki:SPY2}. The shortest period of these is WD0957-666
with $P = 88\,$minutes (Fig.~\ref{fig:wd0957-666})
\cite{Bragaglia:DDs,Maxted:DDmassratios,Moran:WD0957-666}.
There are estimated to be of order 100 million such systems in our galaxy, and
they should be steadily spiralling towards shorter periods under the action of
gravitational wave radiation. The existence of systems such as WD0957-666, which
will merge within 200 million years, proves that there must be a
population of much shorter period systems that we have yet to detect. This was
realised some time ago and it is thought that this population, along with their
accreting counterparts, the AM~CVn stars, will be the dominant gravitational
wave emission sources at frequencies of order $10^{-3}\,\mathrm{Hz}$, right in
the waveband of the Laser Interferometer Space Antenna (\emph{LISA})
\cite{Hils:GWR,Nelemans:GWR0,Nelemans:GWR} which unlike ground-based detectors
is sensitive to gravitational waves of relatively long period . \emph{LISA} is predicted to be able
to detect several thousand of these sources and will be able to narrow down the
location of many of them to less than one degree \cite{Cooray:DWDs,Stoeer:DWDs}, and it should be possible to
optically identify some of them. Optical follow-up can help in the determination
of parameters from the \emph{LISA} data, but will not be easy: the majority of
sources will be fainter than $V = 22$, and yet have periods of order 10 minutes.
Obtaining spectroscopy of these will clearly require very low noise detectors,
even on the largest telescopes, including ELTs. For instance, when observing
the $88\,$ minute period double white dwarf WD0957-666 \cite{Moran:WD0957-666},on the
$3.9\,$m Anglo-Australian Telescope, our exposure times of 500 seconds were a
compromise between smearing the spectra over too much of the orbit versus
suffering too much readout noise, and this was on a relatively bright target
with $V = 14.6$. This is an explicit example of the constraints discussed for Doppler
tomography in section~\ref{subsec:accretion}.

A similarly challenging application for high-speed spectroscopy emerges from the
eclipses of white dwarfs in detached white dwarf/main-sequence binaries. These
are relatively easily studied photometrically \cite{Brinkworth:NNSer}, but the
extra light losses and dispersion of spectroscopy make them a much tougher
subject for spectroscopy and I am not aware of any spectroscopic studies which
target the eclipses in these objects. There is nevertheless a compelling reason
to do so which is that spectroscopy of the white dwarf as it goes in and out of
eclipse has the potential to determine its rotation rate by measurement of the
radial velocity shifts induced by the eclipse. The rotation rate of white dwarfs
in such systems has implications for the stability of double white dwarf binary
stars \cite{Marsh:mdot}. This requires taking spectra
every 5 seconds or so (preferably less) and once again, for typical systems,
telescopes and instruments, moves us into the realm of readout noise.

This is an appropriate point to change topic and look at the technical
difficulties of high-speed CCD spectroscopy.

\section{CCD spectroscopy}
\label{sec:ccds}

\subsection{Standard CCDs}
There are two key issues which make the use of standard CCDs difficult for high-speed
work. First of all CCDs are usually slow to read out. It is not unusual for CCDs
to be read out at $\sim 100\,$kHz pixel rates, so that an entire chip can take
several tens of seconds to be read out. Many chips can be read out faster, but
then one suffers worse read noise, which can more than offset any advantage of
high-speed. The amount of telescope time spent reading out CCDs is potentially
frightening: consider a telescope which spends 8 hours per night, 365 night/year
devoted to taking spectra of 8 one hour-long spectra per night together with
calibration arcs before and after each one. If each spectrum takes 1 minute to
read out, then \emph{18 solid nights} would be spent reading out the CCDs. Worse
still, any programme that required exposure times shorter than 60 seconds would
run at below 50\% efficiency. As some of the examples of
Sect.~\ref{sec:science} showed, this is not as rare as one might imagine.
Luckily there are ways around slow readouts. First one can group pixels (bin)
and read out sub-sections of the chip (window). These are both often very
effective. More radical is to use a frame transfer CCD allowing one half of the
CCD to be read out while the other is being exposed. The EEV 47-20 CCDs used by
ULTRACAM (Dhillon, this volume) allow most observations to be carried out with a
deadtime of only 24 milliseconds using this technique. This would be more than adequate
for the vast majority of feasible spectroscopic applications.

This brings us to the more fundamental issue of readout noise. This plays a
much more important role in optical spectroscopy than it does in photometry.
The key quantity to have in mind is the variance $V$ on a given pixel which
is given by
\begin{equation}
V = R^2 + C, \label{eq:normalccd}
\end{equation}
where $R$ is the RMS readout noise in electrons (typically $\sim 3$) and $C$ is
the number of electrons in the pixel, and is equal to the number of photons
detected (which in general is the sum of target flux, sky background and dark
counts, although the latter can usually be neglected). This is the simplest
possible version of this relation and assumes perfect flat-fielding, but this is
more often than not a reasonable approximation in the case of high-speed work.
Once $C$ drops below $R^2 \sim 9$, one is starting to lose out significantly to
readout noise. There are techniques to alleviate this: binning again is
important. If one is observing a point source, then it makes no sense to
over-resolve the spatial profile, and in fact under-sampling of the spatial
profile is not always much a drawback except for spotting cosmic rays, therefore
binning in the spatial direction is often very useful. I have found that people
often do not realise how significant an improvement binning spatially can make,
but it not hard to demonstrate. Consider a case where two pixels with a total
count $C = R^2$ are binned into one, and assume that $C$ is dominated by the
object (as opposed to sky background). Then the ratio of signal-to-noises, binned to unbinned, is $(3/2)^{1/2}$,
equivalent to a 50\% increase of exposure time. Apart from this, the only other
option may be to increase the exposure time: consider again the marginal case
with $C = R^2$ when one decides to double the exposure time. Then it is
straightforward to show that the improvement in signal-to-noise corresponds to
the improvement that would be obtained with an $8/3 \sim 2.7$-fold 
rather than simply two-fold increase in exposure
time in the zero readout (Poisson limited) case. Of course increasing the
exposure time may not be an option; after all, one is trying to resolve intrinsic
variability in a target. The only other capability one then has to affect the
final signal-to-noise is not to make things worse by poor reduction; it is well
worth using extraction techniques designed to optimise the signal-to-noise ratio
\cite{Horne:optimal,Marsh:optimal} if at all possible.

\subsubsection{When is one readout noise limited?}
\begin{figure}
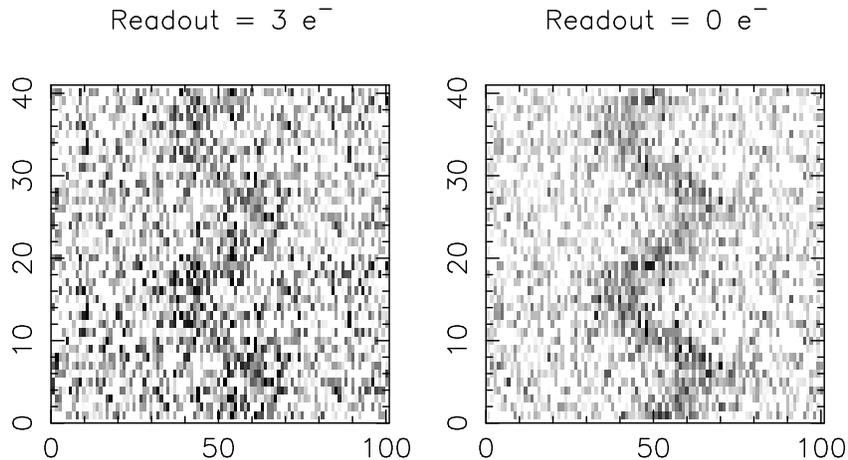

\centering
\hspace*{\fill}
\includegraphics[width=0.45\textwidth]{rxj_sim_3.ps}
\hspace*{\fill}
\includegraphics[width=0.45\textwidth]{rxj_sim_0.ps}
\hspace*{\fill}
\caption{A simulation of the trailed spectrum of the $V = 21$ ultra-compact binary
  star, RX~J0806+1527, as taken with 1 night using the VLT and the FORS2
  spectrograph (600 grism) with (left) and without (right) readout noise. 
  Time advances upwards; the $x$-range is centred on a single emission line.
  The exposure time was taken to be 30 seconds in
  order to resolve the presumed 321 second orbital period of the object, but
  detector readout time was assumed to be zero in each case. The emission line
  was taken to be about as strong as seen in the average spectrum of
  \protect\cite{Israel:RXJ0806}
}
\label{fig:rxj_sim}
\end{figure}

Despite the various ways in which one can combat readout noise, there is in the 
end no getting around it apart from changing the detector design which I look
at next. Before doing so I pause to consider the parameter space where readout
noise is important. Consider a telescope of diameter $D$, feeding a
spectrograph of resolution $R_\lambda$ leading to a detector with $N_d$ pixels per
resolution element in the dispersion direction and $N_s$ pixels across the
spatial profile (loose definitions, but it is the orders of
magnitude that matter here). If $\epsilon$ is the total
throughput of atmosphere, instrument and detector in terms of photons
detected versus those actually incident upon the atmosphere, then an exposure of
$t$ seconds of a target of AB magnitude $m$ will produce $C$ counts given by
\begin{equation}
C = \frac{11853 \epsilon}{N_d N_s R_\lambda} \left(\frac{D}{1\,\mathrm{m}}\right)^2
  \left(\frac{t}{1\,\mathrm{s}}\right) 10^{(16.4-m)/2.5} . \label{eq:divide}
\end{equation}
Assuming $N_d = N_s = 3$, $R_\lambda = 2000$, $D = 8\,\mathrm{m}$, $\epsilon =
0.2$, then to have $C > R^2$ for $R = 3$ electrons implies the following relation between $m$ and $t$:
 \begin{equation}
m < 16.33 - 2.5 \log \left(\frac{t}{1\,\mathrm{s}}\right) .
\end{equation}
One is therefore readout noise limited for a target of $20^\mathrm{th}$ magnitude target and $30$ second
exposures on a typical spectrograph on (currently) the world's largest telescopes. 
Moving to an echelle with, say, $R_\lambda = 60$,$000$ and $\epsilon = 0.1$, then this
limit must be shifted up by $4.5$ mags. A simulation of the effect of readout noise
is shown in Fig.~\ref{fig:rxj_sim}
for a star of considerable current interest, RX~J0806+1527. This is potentially
the shortest period binary known, and a strong gravitational wave source,
consisting of two white dwarfs with an orbital period of just $321$ seconds
\cite{Ramsay:RXJ0806,Israel:RXJ0806}. This has
yet to be proven however, and spectroscopy is by far the most 
promising avenue for testing it. It is however, as
Fig.~\ref{fig:rxj_sim} shows, pushing the capabilities of even the VLT. The fact
that as of mid-2006 this observation had yet to be performed is testament to its 
difficulty. Many would consider $V = 21$ to be ``bright'', or at least, not
especially faint, but it certainly is when one is forced to take short exposures.
Observations of this sort are exactly those needed for the gravitational wave
sources, Sect.~\ref{subsec:whitedwarf}, except they will in general be more
difficult still. 

Having seen the limitations imposed by readout noise, I now turn to how it
can potentially be avoided.

\subsection{Electron-multiplying CCDs}
The readout noise that so seriously limits normal CCDs is added by the
amplifier which converts the small charge on each pixel into a voltage. To
some extent readout noise can be reduced by taking longer over the
double-correlated sampling used in low-noise CCD readouts (i.e. spending
longer integrating the voltage levels before and after the charge on each
pixel is cleared), but $1/f$ noise limits the extent to which one can push
this and one cannot in any case take too long reading each pixel if one is
interested in high speed. Electron-multiplying CCDs (hereafter EMCCDs, but
also known as ``low light level'' or L3CCDs) get round this limitation in a
clever way. In these CCDs, an extra series of stages is added to the serial
register \emph{before} the charge reaches the amplifier. These stages can be
clocked with higher than normal voltages which creates a significant
probability that one electron will generate another as the charges are moved
from stage to stage. With enough stages, a single detected electron can lead
to an avalanche of several hundred or even thousands of electrons. These then
dwarf the readout noise added by the amplifier. The formulae for
signal-to-noise from these devices are more complex than for standard CCDs
\cite{Basden:L3CCD}, but it is necessary to review them here in order to
understand the advantages and limitations of these devices.

To first order, the avalanche gain register can be modelled as a
series of stages at each of which there is a probability $p$ that any
electron will spawn another. If there are $N$ such stages in total
then the mean gain $g$, can be shown to be\footnote{Do not confuse this
  gain with the usual ``gain'' of CCDs which is simply a conversion
  factor between electrons and recorded counts or ADU. To avoid
  confusion I only ever talk here in terms of electrons not ADU.} 
\begin{equation}
g = (1+p)^N .
\end{equation}
The value of $p$, and therefore $g$ is a parameter that can be controlled by
adjustment of the driving voltages. For example if we take $N = 536$
(appropriate to the CCD97 manufactured by the company, e2v) and $p = 0.011$,
then $g = 352$. This gain is itself variable however (after it all it can in
principle lie anywhere from 1 to $2^N \gg g$ although the extremes are
unlikely) which increases the variance of the output over that expected from
pure Poisson noise. Thus in a normal CCD readout, a mean detection rate of $C$
electrons leads to the variance contribution of $C$ in (\ref{eq:normalccd}),
while in an EMCCD the signal is amplified to $g C$ on average and the
corresponding variance is $(g^2 + \sigma^2) C$ where $\sigma^2$ is the
variance of the gain and is given by
\begin{equation}
\sigma^2 = \frac{1-p}{1+p} \left(g^2 - g\right) .
\end{equation}
Had there been no dispersion in the gain then the variance would have
been $g^2 C$, exactly what one would expect if simply
multiplying $C$ by a constant; the variance in the gain thus adss extra noise. 
If, as is the case in practice, $p \ll 1$ and $g \gg
1$, then $\sigma^2 \approx g^2$, and the variance is a factor of $2$
larger than a constant gain would have produced. Thus
(\ref{eq:normalccd}) is modified to
\begin{equation}
V = R^2 + 2 g^2 C, \label{eq:emccd}
\end{equation}
and the signal-to-noise ratio for one pixel is given by
\begin{equation}
\frac{C}{\sqrt{(R/g)^2 + 2 C}},
\end{equation}
compared to
\begin{equation}
\frac{C}{\sqrt{R^2 + C}},
\end{equation}
for a normal CCD. The extra variance can be thought of as being
equivalent to a 50\% drop in QE. Put this way it sounds bad, but
comparing the above two relations for signal-to-noise, and assuming
that $g$ is so large that $R/g$ can be neglected, one can see that the
EMCCD gains once $C < R^2$. In other words once $C \sim R^2$, normal
CCDs have also effectively lost a factor 2 in QE, but unlike EMCCDs
normal CCDs carry on getting worse as $C$ drops. This analysis also
shows that (\ref{eq:divide}) marks the dividing line between normal
CCDs versus EMCCDs running in ``linear'' mode with counts proportional
to the voltage of the amplifier.

At very low count rates, the full QE can be recovered by operating
in a photon counting mode. In this case rather than take the output
divided by the mean gain as an estimate of the counts, one defines a
threshold $T$ such that an output $> T$ is recorded as one photon,
while an output $< T$ is recorded as zero. Provided that one is operating in
a regime where the chance of more than 1 e$^-$/pixel/exposure is
small, this adds no extra noise to the output, and the signal-to-noise
becomes
\begin{equation}
\frac{C}{\sqrt{(R/g)^2 + C}}.
\end{equation}
If $g$ is large, this implies ideal, Poisson-limited performance. Such
a device could give us data looking like the right- rather than the
left-hand side of Fig.~\ref{fig:rxj_sim}.

\subsubsection{Problems: thresholds, non-linearity and Clock-Induced Charges}
\begin{figure}
\centering
\includegraphics[angle=270,width=0.8\textwidth]{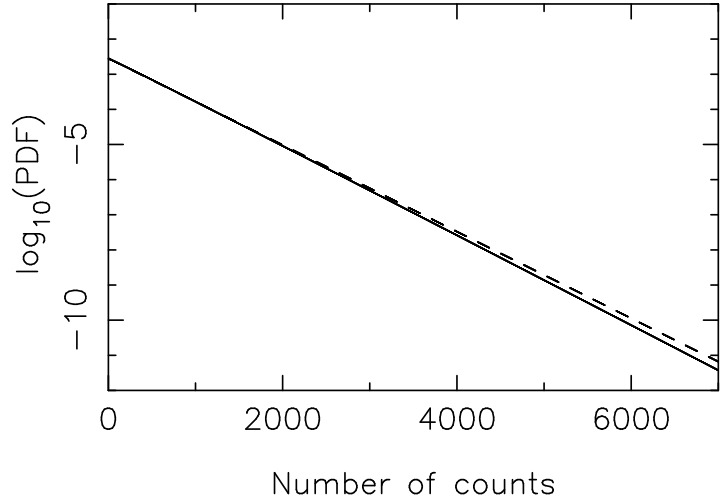}
\caption{The solid line shows an exact computation of the output probability distribution function of a 536 stage
  electron multiplying register with $p = 0.011$ given an input of 1
  electron. The dashed line shows the approximation $\exp(-n/g)/g$
  from \protect\cite{Basden:L3CCD} where $n$ is the number of counts
  and the mean gain $g = 352$ in this case. The approximation only
  deviates significantly for values which are very rare 
}
\label{fig:l3pdf}
\end{figure}

EMCCDs suffer from some drawbacks, that apply to normal CCDs but which are not
usually apparent above readout noise. First of all, given the variable gain,
which can in principle be as low as $1$, a finite threshold must imply some
loss of counts and therefore QE. To know how much, one needs to know the full
probability distribution of the gain. For a single electron input, low $p$,
high $g$ case, the output probability distribution function (PDF) for $n$
counts is quite well approximated by $g^{-1} \exp -n/g$ \cite{Basden:L3CCD}.
Figure~\ref{fig:l3pdf}
shows a comparison between an exact calculation of the PDF and this
simple approximation, showing it to be good. This immediately implies
that a threshold $T$ leads to a decrease in the QE by factor of 
$\exp( -T/g)$. One therefore wants $T \ll g$ to avoid too large a loss
in QE, but at the same time $T$ must not be so low that readout noise
alone leads to spurious counts, i.e. $T > 4$ -- $5$ times $R$, so that for
gaussian noise there is a very small chance of counts induced artificially by
readout noise.

The next problem is a common feature of photon counting devices, which
is non-linearity at ``high'' count rates. In this case significant
non-linearity will set in when the mean count rate rises above $\sim 0.2\,$e/pixel/exposure.
Thus one can be limited by the rate at which pixels can be clocked
out. This is exactly like the IPCS except that one can elect in the
case of EMCCDs to work in the linear mode if it is clear that the
count rates are too high for reliable photon counting.

The third and worst problem is a new phenomenon, or at least one that is
only revealed by the low noise capabilities of EMCCDs: ``Clock Induced
Charges'' or CICs. These are electrons which are spontaneously
produced during clocking. Their statistics are complex and I leave a detailed
discussion of them to the appendix where I include calculations that as far as I
am aware have not been published before. As far as we are concerned they are
equivalent to a readout noise and lead to a variance (in the photon counting
case) of the form
\begin{equation}
V = \left(\frac{R}{g}\right)^2 + R_C + C,
\end{equation}
where $R_C$ is the number of CICs/pixel/exposure, equivalent to a readout noise
before avalanche amplification of $R_C^{1/2}$. The $C$ here should be interpreted as including the $\exp( -T/g)$
factor discussed above, and the $R_C$ rate will similarly be threshold dependent
in this case. The CICs are better thought of as a source of readout noise than
background counts because they are incurred per exposure, not per unit time.

\begin{figure}
\centering
\includegraphics[angle=270,width=0.8\textwidth]{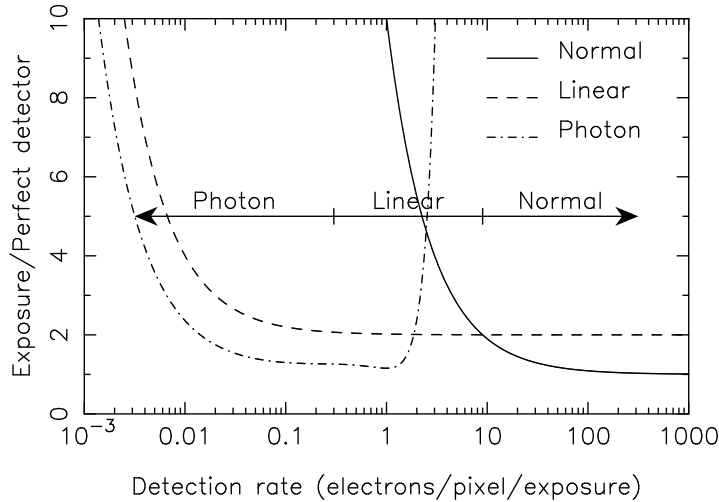}
\caption{The three operating modes of CCDs are shown in terms of the exposure
  time taken to reach a given signal-to-noise compared to a ``perfect'' detector
  of identical QE. Normal mode (solid line) is best at high count levels drops
  off sharply once readout noise dominates. Photon  counting mode (dot-dashed
  line) is best at very low count levels but cannot go past 1
  e$^-$/pixel/exposure owing to non-linearity. In between, linear or
  ``proportional'' mode (dashed line) is best despite the 50\% QE drop caused by
the variable avalanche gain. The rise at very low counts is caused by CICs,
with a rate of $0.01\,$e/pixel/exposure assumed here
}
\label{fig:three_modes}
\end{figure}

The three possible operating modes of CCDs are summarised in
Fig.~\ref{fig:three_modes}.
For EMCCDs the focus is often on photon counting mode and the need to clock the
pixels out fast to retain linearity. Figure~\ref{fig:three_modes} shows that there
remains  a role for the linear mode, the fraction of parameter space it occupies 
depending upon the CIC rate. Note that I have extended the regime of the
linear mode in Fig.~\ref{fig:three_modes} towards lower counts than might seem
justified by the curves; this is because of the non-linearity of photon
counting. In principle it can be corrected to some extent, and this is built
into Fig.~\ref{fig:three_modes}, however in practice non-linearity corrections
do not work as well as one might hope and I have restricted the photon counting
mode to rates less than $0.3$ e$^-$/pixel/exposure.

Clearly the CIC rate is of central importance in how well these devices will
perform in practice, as has been recognised before \cite{Daigle:L3CCD}.  Rates
in the range $0.004$ to $0.1$ CICs/pixel/exposure have been quoted
\cite{Tulloch:L3CCD}.  One can hope for manufacture-driven improvements with
time, and controller improvements have a significant role to play too
\cite{MacKay:L3CCD}, but one should not lose sight of the fact that already even
the high rate of $0.1$ CICs/pixel/exposure is equivalent to an extremely low
readout noise and can allow the count rate $C$ to drop by a factor of 100
compared to normal CCDs before the noise floor is hit; this is an impressive
potential gain and exactly what it needed to carry out high-time resolution
spectroscopy on faint objects.  Having said that, CICs do negate the QE
advantages of CCDs compared to alternatives, in particular GaAs image
intensifiers \cite{Daigle:L3CCD}.

\subsubsection{OPTICON-funded L3CCD for spectroscopy}
Given the uncertainties over CIC rates, thresholds, pixel clocking rates and
the like, EMCCDs are very much in the development phase for astronomy. We
don't yet know indeed whether they are reliable for spectroscopy in the sense
that they can return accurate atomic line profiles, equivalent widths and
radial velocities. Thus as part of the OPTICON programme of the EU's Framework
Programme 6 (FP6), the Universities of Sheffield and Warwick, and the
Astronomy Technology Centre, Edinburgh have started a programme to
characterise an EMCCD for spectroscopic work. We have purchased a CCD201 which
is a frame transfer device with a 1k~$\times$~1k imaging area and with one
normal and one avalanche readout.  As of mid-2006, the device is mounted in a
cryostat and producing test data. We will be using hardware developed for the
high-speed camera ULTRACAM \cite{Dhillon:ULTRACAM}. In terms of imaging area,
this device is not competitive with existing detectors on spectrographs. On
top of this the ULTRACAM controller will not push the device to its limits and
so clear improvements are possible even now. However, it will allow us to see
whether these devices are a promising route to explore for future development
of high-speed spectroscopy.

\section{Conclusion}

High-speed optical spectroscopy offers a number of unique diagnostics of
extreme astrophysical environments and will become only more necessary as
larger surveys discover more unusual objects. Unfortunately standard CCDs, the
workhorses of optical astronomy, are not well suited to the high-speed and low
noise required when light is dispersed. This may be changing with recent
developments of avalanche gain CCDs, but they must be tested on celestial
objects before we know whether this is truly the case. This is the aim of a
programme funded under the EU FP6's OPTICON consortium.

\section*{Acknowledgements}
I thank PPARC and OPTICON for funding during this work, and Vik Dhillon, Andy
Vick, Derek Ives and Dave Atkinson for their help and
collaboration in this work.

\bibliographystyle{plain}
\bibliography{refs}

\appendix
\section{Clock Induced Charge statistics}
\begin{figure}
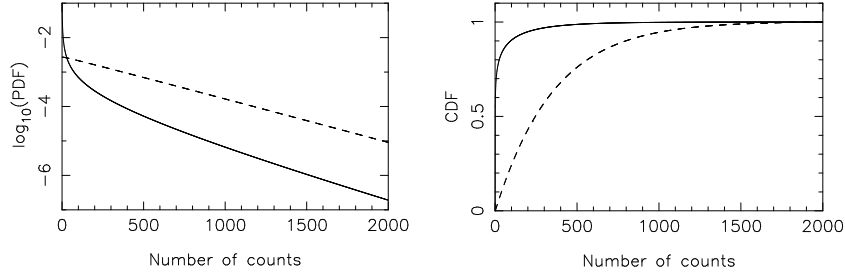

\centering
\hspace*{\fill}
\includegraphics[angle=270,width=0.45\textwidth]{cic.ps}
\hspace*{\fill}
\includegraphics[angle=270,width=0.45\textwidth]{ciccdf.ps}
\hspace*{\fill}
\caption{In the left panel the solid line shows the distribution of in-register CICs for a 536 stage
  electron multiplying register with $p = 0.011$ and CIC probability/stage of $p_c = 0.0011$, i.e. this
  is the output PDF given a zero electron input. The dashed line shows the PDF
  given an input of 1 electron with $p_c$ set to zero. The right-hand panel
  shows the equivalent cumulative distribution functions, and illustrates that
  the parameters chosen here will give a CIC above 100 counts in about 1 in 10 pixels
}
\label{fig:cic}
\end{figure}

In this section we detail some of the statistical properties of CICs.
CICs come in two varieties: ``pre-register'' and ``in-register'' events.
Pre-register events suffer the full amplification of the avalanche stage
and will have identical statistics to electrons generated from genuine signals.
In-register events are amplified by a variable amount depending upon where in
the avalanche register they are first produced. This leads to an extremely
skewed distribution for these events with lots of low values but a tail
extending to very high values as well as shown in Fig.~\ref{fig:cic}.
The skewed distribution makes it hard to define CIC rates in a general way.
For example, Fig.~\ref{fig:cic} shows that for the particular parameters
chosen about 25\% of pre-register CICs would on output fall below a threshold
of 100, while 90\% of the in-register CICs will do so. Lowering the threshold
could increase the pre-register rate by at most 25\%, but could increase the
in-register rate by up to 10 times. When quoting rates, it is important to
define how they were measured. In this case a threshold of 100 will lead to a
rate of $0.1$ CIC/pixel/exposure, at the high end of total CIC rates (in- plus
pre-register) \cite{Tulloch:L3CCD}, and so presumably 
the probability of a CIC being generated at any one stage of the avalanche
register is usually less than I have assumed.

If the probability of a CIC being generated at any one stage of the avalanche
register is $p_C$, then one can show that the mean output value, given zero
input $\mu_C$ is given by
\begin{equation}
\mu_C = \frac{p_C}{p} (g - 1),
\end{equation}
while the variance is
\begin{equation}
\sigma^2_C = \left( \frac{2/(1+p) - p_c}{2 + p} (g + 1) - \frac{1-p}{1+p}\right) \mu_C .
\end{equation}
For $p \ll 1$, $p_C \ll 1$, $g \gg 1$, this boils down to $V_C \approx g \mu_C$.
For the example shown in Fig.~\ref{fig:cic}, $\mu_C = 35$ while $\sigma_C =
110$. In proportional mode, this is effectively equivalent to a readout noise
component of $R = \sigma_C/g = 0.3$ entering the variance equation $V = R^2 + 2
C$ (ignoring the amplifier readout noise). A threshold of 100 leads to a very similar
number for photon counting mode. 

There may be some room for optimisation of the threshold in the presence of 
significant numbers of in-register CICs. In the
example shown, a threshold of 100 leads to a 25\% reduction in the true event
rate and a 10\% CIC rate, so that the signal-to-noise would be 
\begin{equation}
\frac{0.75 P}{\sqrt{0.1 + 0.75 P}},
\end{equation}
where $P$ is the mean number of photo-electrons per pixel per exposure. If the
threshold is raised to 200, this becomes
\begin{equation}
\frac{0.57 P}{\sqrt{0.05 + 0.57 P}},
\end{equation}
which is better for $P < 0.057$. 

The threshold cannot be optimised for pre-register events in the same way since they have
identical statistics to genuine events. The relative numbers of in-register and
pre-register events appears to depend upon the controller
\cite{Tulloch:L3CCD,MacKay:L3CCD}. My suspicion, which is possibly borne out by
one of these studies \cite{MacKay:L3CCD}, would be that the high voltages required for the
avalanche register will make it harder to reduce the in-register compared to
pre-register event rates.



\printindex
\end{document}